# Expanding Space:
# The Root of Conceptual Problems of the Cosmological Physics

© **Yu. Baryshev**[1,2]

[1] Astronomical Institute of the St.-Petersburg State University, St.-Petersburg, Russia
[2] Email: yuba@astro.spbu.ru

**Abstract:** The space expansion physics contains several paradoxes which were clearly demonstrated by Edward Harrison (1981, 1995, 2000), who emphasized that the cooling of homogeneous hot gas (including photon gas of CBR) in the standard cosmological model based on the violation of energy conservation by the expanding space. In modern version of SCM the term ``space expansion'' actually means continuous creation of vacuum, something that leads to conceptual problems. Recent discussion by Francis, Barnes, James, and Lewis (2007) on the physical sense of the increasing distance to a receding galaxy without motion of the galaxy is just a particular consequence of the arising paradoxes. Here we present an analysis of the following conceptual problems of the SCM: the violation of energy conservation for local comoving volumes, the exact Newtonian form of the Friedmann equation, the absence of an upper limit on the receding velocity of galaxies which can be greater than the speed of light, and the presence of the linear Hubble law deeply inside inhomogeneous galaxy distribution. The common cause of these paradoxes is the geometrical description of gravity, where there is no a well defined concept of the energy-momentum tensor for the gravitational field, no energy quanta – gravitons, and no energy-momentum conservation for matter plus gravity because gravity is not a material field.

## 1. Space expansion paradigm in general relativity

*Einstein's equations and Bianchi identity.*

The basic element of the standard cosmological model (SCM) is the general relativity (GR), which is a non-quantum geometrical gravity theory. Classical relativistic gravity effects were predicted by GR and successfully tested for the weak gravity conditions in the solar system and binary neutron stars. It is assumed that GR can be applied to the Universe as a whole and hence describe cosmological models.

According to GR gravity is described by a metric tensor $g^{ik}$ of a Riemannian space. The "field" equations in GR may be written in the form (we use Landau & Lifshitz 1971 notations):

$$R^{ik} - \frac{1}{2} g^{ik} R = \frac{8\pi G}{c^4}(T_m^{ik} + T_{de}^{ik}) \,, \qquad (1)$$

where $R^{ik}$ is the Ricci tensor, $R$ is the scalar curvature, $T_m^{ik}$ is the energy-momentum tensor (hereafter EMT) for usual matter, and $T_{de}^{ik}$ is the EMT of the dark energy component, which includes the cosmological constant and cosmological vacuum. The most important feature of the Einstein's eq.(1) is that the right part does not include the energy-momentum of the gravity field itself.

From the Bianchi identity one gets the continuity equation in the form:

$$(T^{ik})_{;k} = (T_m^{ik} + T_{de}^{ik})_{;k} = 0 \,, \qquad (2)$$

where $T^{ik}$ is the total EMT of the matter and dark energy and in the case of non-interacting matter and dark energy the covariant divergence of each EMT equals zero separately.

Note that gravity in GR is not a kind of matter, so the total EMT does not contain the EMT of gravity field. This is why eq.(2) is not a conservation law for gravity plus total matter (Landau & Lifshitz 1971 [19], sec.101, p.304), though in external gravity field it could be interpreted as conservation laws. We shall call eq.(2) the continuity equation.

*Einstein's cosmological principle.*

The second basic element of the SCM is the Einstein's Cosmological Principle, which states that the universe is spatially homogeneous and isotropic on "large scales" (see e.g. Weinberg 1972 [31]; Peebles

1993 [23]; Peacock 1999 [22]). Here the term "large scales" relates to the fact that the universe is certainly inhomogeneous at scales of galaxies and clusters of galaxies. Therefore, the hypothesis of homogeneity and isotropy of the matter distribution in space means that starting from certain scale $r_{hom}$, for all scales $r > r_{hom}$ we can consider the total energy density $\varepsilon = \rho c^2$ and the total pressure $p$ as a function of time only, i.e. $\varepsilon(\vec{r},t) = \varepsilon(t)$ and $p(\vec{r},t) = p(t)$. Here the total energy density and the total pressure are the sum of the energy density of ordinary matter and dark energy:

$$\varepsilon = \varepsilon_m + \varepsilon_{de}, \qquad p = p_m + p_{de}. \tag{3}$$

An ideal equation of state is usually considered

$$p = \gamma \rho c^2, \tag{4}$$

where usual matter and dark energy have equations of state:

$$p_m = \beta \varepsilon_m, \quad \text{with} \quad 0 \leq \beta \leq 1 \quad \text{and} \quad p_{de} = w \varepsilon_{de}, \quad \text{with} \quad -1 \leq w < 0.$$

Recently values $w<-1$ also were considered for description the "fantom" energy.

*Space expansion paradigm.*

An important consequence of homogeneity and isotropy is that the line element may be presented in the Robertson-Walker form:

$$ds^2 = c^2 dt^2 - S(t)^2 d\chi^2 - S(t)^2 I_k(\chi)^2 (d\theta^2 + \sin^2\theta\, d\phi^2) \tag{5}$$

where $\chi, \theta, \varphi$ are the "spherical" comoving space coordinates, $t$ is synchronous time coordinate, and $I_k(\chi) = \sin\chi, \chi, \sinh\chi$ corresponding to curvature constant values k = +1,0,-1 respectively, $S(t)$ is the scale factor of the Friedmann model.

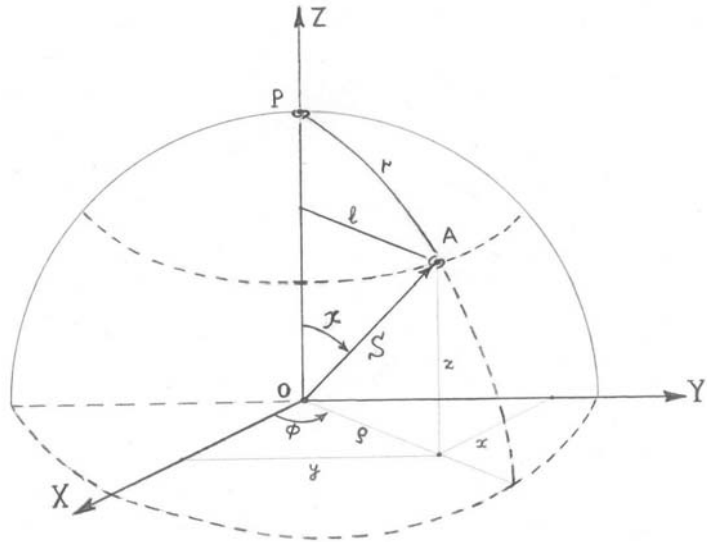

Fig.1. Comoving metric distances: internal $r$, given by eq.(6), and external $l$, given by eq.(8). Radius of the sphere is the scale factor $S(t)$ of the Friedmann model.

The *expanding space paradigm* states that the proper (internal) metric distance $r$ to a galaxy with fixed comoving coordinate $\chi$ from the observer is given by

$$r = S(t) \cdot \chi \tag{6}$$

and increases with time *t* as the scale factor *S(t)*. Note that physical dimension of metric distance [r] = cm , hence if physical dimension [S] = cm, then $\chi$ is the dimensionless comoving coordinate distance. In direct mathematical sense $\chi$ is the spherical angle and *S(t)* is the radius of the sphere (or pseudosphere) embedded in the 4-dimensional Euclidean space.

It means that the "cm" (the measuring rod) itself is defined as unchangeable unit of length in the embedding 4-d Euclidean space. The distance *r* , which is measured in such external units ("cm"), usually is called the "internal" or proper comoving distance on the 3-dimensional hypersurface of the embedding space. In other words *r* and $\chi$ represent the Eulerian and Lagrangian comoving distances.

Often, "cylindrical" comoving space coordinates $\mu, \theta, \varphi$ are used in the literature. In this case the line element is

$$ds^2 = c^2 dt^2 - S(t)^2 \frac{d\mu^2}{1-k\mu^2} - S(t)^2 \mu^2 (d\theta^2 + \sin^2\theta \, d\phi^2) \ , \tag{7}$$

and the comoving "external" metric distance *l* may be introduced as

$$l = S(t) \cdot \mu \ , \tag{8}$$

which mathematically presents the external distance from z-axis in an embedding Euclidean 4-dimensional space. The relation between these two metric distances is given by means of the inversion function for $I_k$ :

$$r = S(t) \cdot I_k^{-1}(l/S) \ , \tag{9}$$

*Exact velocity-distance law in Friedmann's models.*

It is important to point out that the hypothesis of homogeneity and isotropy of space implies that for a given galaxy the recession velocity is proportional to distance. Indeed, due to the "space expansion" the metric distance between any two galaxies becomes larger and the recession velocity $V_{exp}$ of a body with fixed Lagrangian comoving coordinate $\chi$ is the rate of increasing of the metric distance *r* as a function of time. The exact general relativity expression for the expansion velocity is the time derivative of the metric distance given by eq.6, so that

$$V_{exp} = \frac{dr}{dt} = \frac{dS}{dt} \chi = \frac{dS}{dt} \frac{r}{S} = H(t)\, r = c \frac{r}{r_H} \ , \tag{10}$$

where $H(t) = \dot{S}/S$ is the Hubble constant (also is a function of time) and $r_H = c/H(t)$ is the Hubble distance at the time *t*. Here and in the following the dot indicates derivative with respect to the cosmic synchronous time *d/dt*. This means that the theoretically predicted linear velocity-distance relation *V=Hr*, identified with the observed redshift-distance Hubble law, can exist only if the matter distribution is uniform: "the connection between homogeneity and Hubble's law was the first success of the expanding world model" (Peebles et al.,1991 [24]). However as we shall discuss later, according to modern data on galaxy distribution, this is not to be the case at least for luminous matter.

*Friedmann's equations.*

In comoving coordinates the total EMT has the form:

$$T_k^i = diag(\varepsilon, -p, -p, -p) \ , \tag{11}$$

and for the case of unbounded homogeneous matter distribution the Einstein's equations (eq.1) are directly reduced to the Friedmann's equations (FLRW – model). From the initial set of 16 equations we have only two independent equations for the (0,0) and (1,1) components, which may be written in the form:

$$\frac{\dot{S}^2}{S^2} + \frac{kc^2}{S^2} = \frac{8\pi G}{3c^2}\varepsilon \quad, \tag{12}$$

$$2\frac{\ddot{S}}{S} + \frac{\dot{S}^2}{S^2} + \frac{kc^2}{S^2} = -\frac{8\pi G}{c^2}p \quad. \tag{13}$$

From the Bianchi identity (eq.2), the continuity equation takes the form:

$$3\frac{\dot{S}}{S} = -\frac{\dot{\varepsilon}}{\varepsilon + p} \quad, \tag{14}$$

which must be added to eqs.(12, 13) as the consistency condition. Using the definition of the Hubble constant $H = \dot{S}/S$, we rewrite eq.(12) as

$$H^2 - \frac{8\pi G}{3}\rho = -\frac{kc^2}{S^2} \quad, \tag{15}$$

and equation (13) as

$$\ddot{S} = -\frac{4\pi G}{3}(\rho + \frac{3p}{c^2})S \quad. \tag{16}$$

In terms of the critical density $\rho_{crit} = 3H^2/8\pi G$, the total density parameter $\Omega = \rho/\rho_{crit}$, the curvature density parameter $\Omega_k = kc^2/S^2H^2$, and the deceleration parameter $q = -\ddot{S}S/\dot{S}^2$, these equations also may be presented in the form:

$$1 - \Omega = -\Omega_k \quad, \tag{17}$$

$$q = \frac{1}{2}\Omega(1 + \frac{3p}{\rho c^2}) \quad, \tag{18}$$

were *Ω, p, ρ* are the total quantities, i.e. the sum of corresponding components for matter and dark energy. Solving the Friedmann's equation (18) one finds the dependence on time the scale factor *S(t)* or the metric distance *r(t)*, which is the mathematical presentation of the space expansion.

*What does space expansion mean physically?*

Each comoving finite box in expanding universe continuously increases its volume, so gets more and more cubic centimeters. Physically expansion of the universe means the creation of space together with physical vacuum. Creation of space may be visualized by 2-d analogy with expanding sphere in 3-d space, where the surface of the sphere increases with time and for 2-d beings their universe grows with time (gets more square centimeters) .

Real Universe is not homogeneous, it contains atoms, planets, stars, galaxies. Bondi (1947) considered spherical inhomogeneities in the framework of GR and showed that inside them the space expands slowly. In fact bounded physical objects like particles, atoms, stars and galaxies do not expands. So inside these objects there is no space creation. This is why the creation of space is a new cosmological phenomenon, which cannot be tested in laboratory because the Earth, the Solar System and the Galaxy do not expand.

Below we consider several puzzling properties of the expanding space, which are a direct consequence of the above derived exact equations.

## 2. Violation of the energy conservation in expanding space

Intriguingly the continuity equation (14) can be written also in the form

$$dE + p\, dV = 0 \; , \tag{19}$$

where $dE = d(\varepsilon V) = d(\rho c^2 V)$ is the change of energy within the comoving volume $V = const \times S^3$ V.
Eq.(19) looks like the law of conservation of energy in the lab thermodynamics.

However, as it was emphasized by Harrison (1981, 1995) there is an essential difference between the lab and the cosmological cases. Eq.(19) in the laboratory reads: if inside a finite box the energy decreases, then it reappears outside the box as the work produced by the pressure (e.g. acting on a piston of a machine, increasing the volume of the box). The work performed by the pressure inside the box is the cause of the energy decrease inside the box.

Contrary to the lab case, in expanding space the cosmological pressure does not produce work. It was noted by Harrison(1981; 1995) that in a homogeneous unbounded expanding FLRW model one may imagine the whole universe partitioned into macroscopic cells, each of comoving volume $V$, and all having contents in identical states. The $-pdV$ energy lost from any one cell cannot reappear in neighboring cells because all cells experience identical losses. So the usual idea of an expanding cell performing work on it surroundings cannot be applied to the cosmological case.

In cosmology eq.(19) gives us a possibility to calculate of how much the energy increases or decreases inside a finite comoving volume but it does not tell us where the energy comes from or where it goes. As Edward Harrison emphasized: "The conclusion, whether we like it or not, is obvious: energy in the universe is not conserved" (Harrison, 1981, p.276). The same conclusion was reached by Peebles (1993) when he considered the energy loss inside a comoving ball of the photon gas. On page 139 he wrote "The resolution of this apparent paradox is that while energy conservation is a good local concept, ... there is not a general global energy conservation in general relativity." But what is more there is no also local energy conservation in each comoving cell, and the root of the puzzle is in the geometrical description of the gravity.

As Landau & Lifshitz (1971) emphasized in paragraph 101 ("The energy-momentum pseudotensor") the expression for Bianchi identity (eq.2) has the form

$$(T^{ik})_{;k} = \frac{1}{\sqrt{-g}} \frac{\partial(\sqrt{-g}\, T^{ik})}{\partial x^k} - \frac{1}{2} \frac{\partial g_{kl}}{\partial x_i} T^{kl} = 0 \; , \tag{20}$$

and because of the mathematical structure of the covariant divergence in Riemannian space the eq.20 has an extra (second) term which violate the integral conservation $\int T^{ik} \sqrt{-g}\, dS_k$, for which the condition $\partial(\sqrt{-g}\, T^{ik})/\partial x^k = 0$ should be fulfilled. This is why according to Landau & Lifshitz the equation (20) does not generally express a law of conservation. To get the total (all kinds of matter plus gravity) energy-momentum conserved, they suggest to consider energy-momentum pseudotensor, which could describe energy density of the gravity field itself. However this violates the tensor character of the laws of conservation. The root of the problem lies in the equivalence principle and in the absence of a true gravity force in GR, while all other fundamental fields have true forces, true EMTs and operate in Minkowski space. It is important that Noether theorem relates a conserved EMT of a material field to maximal symmetry of the Minkowski space and this is why in curved Riemannian space the EMT of gravity field can not be properly defined.

The problem of the absence of a true EMT for gravity field in cosmology appears as the violation of energy conservation during the space expansion. Indeed, let us consider the energy content of a comoving ball with radius $r(t) = S(t)\, \chi$ . The volume element in metric eq.(5) is

$$dV = S^3 I_k^2(\chi)\sin(\theta)\,d\chi\,d\theta\,d\phi \quad, \tag{21}$$

and energy content of the comoving sphere is

$$E(r,t) = \int_0^r T_0^0 \, dV = \frac{4\pi}{3}\varepsilon(t)\, S^3(t)\, \chi^3\, \sigma_k(\chi) \quad, \tag{22}$$

where $\sigma_k(\chi) = \int_0^\chi I_k^2(y)\, dy$, so it equals to 1 for k = 0, to $3/\chi^3(\chi^2/2 - \sin 2\chi/4)$ for k = 1, and to $3/\chi^3(\sinh 2\chi/4 - \chi^2/2)$ for k = -1.

To calculate the time dependence of the energy density $\varepsilon(t) = \rho(t)c^2$ we use the continuity equation (14) and an ideal equation of state $p = \gamma\,\rho c^2$. Then the energy inside a comoving ball with radius $r$ will change with time as

$$E(r,t) = \frac{4\pi}{3}\rho(t)c^2\, r^3(t)\,\sigma_k(\chi) \quad \propto \quad S^{-3\gamma}(t) \quad, \tag{23}$$

so that for dust, radiation and vacuum within the comoving sphere of radius $r$ we get

$$E_{dust}(t) \propto const, \qquad E_{rad}(t) \propto S^{-1}(t), \qquad E_{vac}(t) \propto S^{+3}(t), \tag{24}$$

In fact, only for dust (where p=0) one may speak about energy conservation in expanding universe. But for any other matter with $p \neq 0$ within any local comoving volume energy is not conserved. This is because in GR there is no EMT of gravity field itself, which could play the role of an additional substance to restore the conservation laws.

### 3. Newtonian form of the relativistic Friedmann equation

Because of Lagrangian comoving coordinate $\chi$ does not depend on time, the exact Friedmann equation (16) can be also written in the form, where the metric distance $r = S(t) \cdot \chi$ appears explicitly (eq.6), so that

$$\frac{d^2 r}{dt^2} = -\frac{G M_g(r)}{r^2} \quad, \tag{25}$$

where the active gravitating mass $M_g(r)$ of a comoving ball with radius $r$ is given by the exact relation

$$M_g(r) = -\frac{4\pi}{3}(\rho + \frac{3p}{c^2})\, r^3 \quad. \tag{26}$$

Exact Friedmann equation in the form eq.(25) is identical with the Newtonian equation where matter density contains also the term $3p/c^2$. Multiplying eq.25 by the mass of a receding galaxy one gets the cosmological Friedmann force acting on a test galaxy placed at distance $r$ from a fixed point at the center of coordinate system:

$$F_{Fr} = m \frac{d^2 r}{dt^2} = -\frac{G\, m\, M_g(r)}{r^2} \quad . \tag{27}$$

Therefore the exact relativistic equation describing the dynamical evolution of the universe is exactly equivalent to the non-relativistic Newtonian equation of motion of a test particle in the gravity field of a finite sphere containing a mass $M_g$ within the radius $r$. The second term in eq.(26) does not change the Newtonian character of the solutions.

Such a similarity was first mentioned by Milne(1934) [21] and McCrea \& Milne(1934) [20], though they consider the Newtonian model an approximation to Friedmann model. Later many authors claimed that the Newtonian model can be used only for small radius compared to the horizon distance. Here, however, we see that the Newtonian form of the Friedmann equation is exact and true for all radius. This creates a problem in cosmology because eq.(27) places neither such relativistic restrictions as motion velocity less than velocity of light, nor retardation response effect. The root of the puzzle lies in the geometrical description of gravity in GR and in the derivation of Friedmann's equation from Einstein's gravity equations, using the comoving synchronous coordinates with universal cosmic time $t$ and homogeneous unbounded matter distribution.

The Newtonian form of the Friedmann equation also creates the so called Friedmann-Holtsmark paradox. According to the Friedmann equation there is the cosmological force eq.(27) acting on a galaxy situated at the distance $r$ from another fixed galaxy. This is in apparent contradiction with well known Holtsmark result for the probability density of the force acting between particles in infinite Euclidean space in the case of $1/r^2$ behavior of the elementary force (Holtsmark,1919 [16]; Chandrasekhar,1941 [6]). For symmetry reasons, due to the isotropy of the distribution of particles the average force in any given location is equal to zero and one is left with the finite value of fluctuating force, which is determined by the nearest neighbor particles. Hence in infinite Euclidean space with homogeneous Poisson distribution and Newtonian gravity force there is no global expansion or contraction, but there is the density and velocity fluctuations caused by local gravity force fluctuations.

*Continuous creation of gravitating mass*

A puzzling property of the FLRW model also come from consideration of the active gravitating mass of the cosmological fluid, which may be either positive or negative and changes sign with the cosmic time. In the case of one fluid with equation of state $p = \gamma \rho c^2$ the active gravitating mass (eq.26) will be

$$M_g(r) = \frac{4\pi}{3}(1+3\gamma)\,\rho\, r^3 \quad \propto \quad S^{-3\gamma}(t) \quad . \tag{28}$$

Hence for the case of dust the gravitating mass does not depend on time, but in the case of radiation the gravitating mass continuously disappear in the expanding universe. The most strange example is the vacuum, where $\gamma = -1$ and the gravitating mass is negative. This means that vacuum antigravity continuously increases in time due to the continuous creation of gravitating (actually "antigravitating") vacuum mass. In this sense the continuous creation of matter in the Steady State cosmological model is just a particular case of the new physics of the expanding space.

## 5. Lemaitre's effect of cosmological redshift

Harrison (1981; 1993) [13, 14] clearly demonstrated that the cosmological redshift due to the expansion of the universe is a new physical phenomenon and is not the well known in laboratory the Doppler effect. Recently this subject was intensively discussed by Kiang (2003) [18], Davis \& Lineweaver (2003) [8], Whiting (2004) [32] and Abramowicz et al. (2006) [1] in an attempt to clarify some "common big bang misconceptions" and the "expanding confusions" widely spread in the literature. A summary of the discussion was done by Francis et al. (2007) who also cited Rees & Weinberg (1993) state: " … how is it possible for space, which is utterly empty, to expand? How can nothing expand? The answer is: space does not expand. Cosmologists sometimes talk about expanding space, but they should know better." While Harrison (1981) stated: "expansion redshifts are produced by the expansion of space between bodies that are stationary in space".

In the SCM the cosmological redshift is a new physical phenomenon due to the expansion of space, which induce the wave stretching of the traveling photons via the Lemaitre's equation:

$$(1+z) = \lambda_0 / \lambda_1 = S_0 / S_1 , \qquad (29)$$

where $z$ is cosmological redshift, $\lambda_0$, $\lambda_1$ are the wavelengths at the emission and reception, respectively, and $S_0$, $S_1$ are the corresponding values of the scale factor. Equation (29) may be obtained from the radial null-geodesics of the FLRW line element.

The cosmological redshift (29) is caused by the new physical phenomenon, the Lemaitre's effect, which is different from the familiar in lab the Doppler's effect. On can also see this if compare relativistic Doppler and cosmological FLRW velocity-redshift V(z) relations. To get V(z) in SCM one should consider first V(r) and then r(z) relations. Exact velocity-distance relation in FLRW model is given by eq.(10): $V_{\exp}(r) = H(t)\, r$, and the exact distance-redshift $r(z)$ relation in FLRW model is:

$$r(t_0, z) = r(z) = \frac{c}{H_0} \int_0^z \frac{dz'}{h(z')} , \qquad (30)$$

where $h(z)$ is taken from Friedmann equation (15):

$$h(z) = \sqrt{\tilde{\rho}(z)\, \Omega_0 + (1-\Omega_0)(1+z)^2} , \qquad (31)$$

where $\Omega_0 = \rho_{tot}^0 / \rho_{crit}^0$ is the density parameter in present epoch, $\tilde{\rho}(z) = \rho_{tot} / \rho_{tot}^0$ is the normalized density of the total substances. Analytical expressions for $r(z)$ may be obtained only in some simple cases. For example for empty universe FLRW model with $q_0 = 0$ one gets

$$r(z) = \frac{c}{H_0} \frac{z\,(1+\frac{z}{2})}{1+z} . \qquad (32)$$

Now we can compare the exact FLRW $V(z)$ relation with exact relativistic Doppler relation:

$$V_{\exp}(z) = c\, \frac{r(z)}{r_{H_0}} , \qquad \text{and} \qquad V_{Dop} = c\, \frac{2z+z^2}{2+2z+z^2} . \qquad (33)$$

Clearly these are two different mathematical formulae which corresponds to two different physical phenomena -- Lemaitre and Doppler effects. Eqs.(32) give the same results only in the first order of $V/c$, however the physics of space expansion is different from motion in static space. Indeed, the expansion velocity $V_{\exp} > c$ for such redshifts where metric distances $r > r_H$, while the Doppler velocity is less than $c$ for any large $z$.

## 6. Bondi's effect of cosmological gravitational redshift

In 1947 in the classic paper "Spherically symmetrical models in general relativity" by Sir Hermann Bondi it was shown that, at least for small redshifts, the total cosmological redshift of a distant body may be expressed as the sum of two effects: the velocity shift (Doppler effect) due to the relative motion of source and observer, and the global gravitational shift (Einstein effect) due to the difference between the potential energy per unit mass at the source and at the observer. It means that the spectral shift depend on the distribution of matter in the space around the source.

In the case of small distances Bondi derived a simple formula for redshift which is simply the sum of Doppler and gravitation effects, and which explicitly showed that "the sign of the velocity shift depends on the sign of $v$, but the Einstein shift is easily seen to be towards the red" (Bondi,1947 [6],p.421). Hence according to Bondi the cosmological gravitational frequency shift is redshift (contrary to Peacock 1999 [22], p.619 and Zeldovich \& Novikov 1984 [33] p.97 considerations).

It was shown by Baryshev et al.(1994) [2] that from Mattig's distance-redshift relation one can derive directly for the case of $z<<1$, $v/c \approx x = r/r_H$ the relation for cosmological redshift in the form the sum of the Doppler and the gravitational redshifts

$$z_{cos} \approx z_{Dop} + z_{grav} = x + \frac{1+q_0}{2} x^2 = (\frac{v}{c} + \frac{v^2}{2c^2}) + \frac{q_0}{2} x^2 \quad, \tag{34}$$

where the cosmological gravitational redshift is

$$z_{grav} = \frac{\delta \phi(r)}{c^2} = \frac{1}{2} \frac{G M(r)}{c^2 r} = \frac{1}{4} \Omega_0 \, x^2 \quad. \tag{33}$$

Note that the eq.(33) describes the global gravitational shift due to the whole mass within the ball having the center at the source and the radius equal to the distance between the source and the observer. Hence cosmological gravitational shift depends on the whole matter distribution between the source and the observer (and should not be confused with the local gravitational shift at the source).

It is important that the center of the ball is placed at the source. Then the cosmological gravitational redshift is consistent with the causality principle according to which the event of emission of a photon by the source (which marks the centre of the ball) must precede the event of detection of the photon by an observer on the surface of the ball. The detection event marks the spherical edge of the ball, where all potential observers are situated.

In the literature there are a few discussions of the cosmological gravitational shift but they contain mistaken claims. For instance, if one consider the observer at the center of the cosmological ball and a galaxy at the edge of the sphere, then one may conclude that cosmological gravitational shift is blueshift (see Zeldovich & Novikov,1984 [33], p.97). Also one should use proper metric distance for calculation the mass within a ball, instead of angular distance used in Peacock,1999 [22], problem 3.4.

Note that in the case of the fractal matter distribution with fractal dimension D=2 the cosmological gravitational redshift gives the linear distance-redshift relation and becomes an observable cosmological phenomenon (see e.g. Baryshev et al.1994 [2], Baryshev 2008b).

**7. Hubble-deVaucouleurs paradox**

According to SCM the linear Hubble law is a consequence of the homogeneity of the matter distribution. However studies of the 3-dimensional local galaxy Universe have shown that at least in the range of scales 1 - 100 Mpc galaxy distribution is strongly inhomogeneous and has fractal properties (e.g. Sylos Labini et al.,1998 [25]; Baryshev \& Teerikorpi 2006 [5]). This confirms de Vaucouleurs' prescient view on the matter distribution so we call it de Vaucouleurs law of large scale galaxy distribution (Baryshev et al. 1998 [3]; Baryshev \& Teerikorpi 2002 [4]).

At the same time modern observations of the local Hubble flow based on Cepheid distances to local galaxies, Tully-Fisher distances from the KLUN program, and other distance indicators, demonstrate that the linear Hubble law is well established within the Local Volume ($r<10$ Mpc), starting from distances as small as 1 Mpc (see Teerikorpi,1997 [26]; Ekholm et al.,2001 [9]; Karachentsev et al. 2003 [17]; Teerikorpi et al. 2005 [27]).

A puzzling conclusion is that the strictly linear redshift-distance relation is observed just inside inhomogeneous galaxy distribution, i.e. deep inside the fractal structure for distances less than homogeneity scale (it is known that $r_{hom} > 100$ Mpc):

$$( \, r < r_{hom} \, ) \quad \& \quad ( \, cz = H_0 \, r \, )$$

This empirical fact presents a profound challenge to the standard model where the homogeneity is the basic explanation of the Hubble law, and "the connection between homogeneity and Hubble's law was the first success of the expanding world model" (Peebles et al.,1991 [24]). In fact, within the SCM one would not expect any neat relation of proportionality between velocity and distance for close galaxies, which are members of large scale structures. However, contrary to the expectation, modern data show a good linear Hubble law even for nearby galaxies. It leads to a new observationally established puzzling fact that the linear Hubble law is not a consequence of the homogeneity of visible matter, just because the visible matter is distributed inhomogeneously.

The Hubble and de Vaucouleurs laws describe very different aspects of the Universe, but both have in common universality and observer independence. This makes them fundamental cosmological laws and it is important to investigate the consequences of their coexistence at the same length-scales (see Baryshev et al.,1998 [3]; Gromov et al. 2001 [12]; Teerikorpi et al. 2005 [27]).

### 8. Problems for quantum approach to geometrical gravity

A cosmological model is in fact a particular solution of the gravity field equations. This is why the roots of the conceptual problems of modern cosmology considered above actually lie in the theory of gravitation. In fact, all fundamental forces in physics (strong, weak, electromagnetic) have quantum nature, (i.e. there are energy quanta of corresponding fields which carry the energy-momentum of the physical interactions), while GR is a non-quantum theory, which presents the geometrical interpretation of gravitational force (i.e. the curvature of space itself which is not material field in space) and exclude the concept of localizable gravitational energy. This is why the main problem of GR is the absence of the energy of the gravity field or pseudo-tensor character of gravity EMT (Landau & Lifshitz,1971 [19]). Together with GR the energy problem comes to cosmology and is the cause of the conceptual problems of SCM.

Many years of attempts to unify general relativity with quantum physics have little success. Conceptual tensions between quantum mechanics and general relativity first were noted by Wigner in 1957 and have recently again attracted attention (e.g. Alley 1995; Yilmaz 1997; Amelino-Camelia 2000; Chiao 2003). They emphasize that the most pressing problem in present-day theoretical physics is how to unify quantum theory with gravitation, so called "the quantum gravity problem". The standard scheme of quantization applied to general relativity gives a theory that is not renormalizable (i.e. leads to inevitable infinities in physical quantities). Other attempts are based on the string/M theory, canonical/loop quantum gravity, non-commutative geometry. However the difficulties on this way so large that there is still no generally accepted quantum geometrical gravity theory.

Existing partial approaches to quantum geometry predicts violation of the equivalence principle, possible violation of the Lorentz invariance, and time-varying fundamental physical constants at such a level that their detection may be realistic in near future (Amelino-Camelia et al. 2005). However, up to now increasingly strong limits have been derived on variations of fundamental constants (Levshakov et al. 2005). Also first observations of sharp images of a very distant supernova did not confirm the predicted quantum structure of space-time at Planck scales (Ragazzoni et al. 2003). There is also no deflection from the Newtonian gravity law at small distances down to μm scales (Nesvizhevsky \& Protasov 2004).

### 9. Conclusion.

It is possible that in cosmology we have an example of a new physical phenomena where conservation laws are violated, receding velocities of whole galaxies may exceed the velocity of light and cosmological redshift is due to space expansion. Note that the explanation of the cooling of the photon gas in SCM, and hence the origin of the cosmic microwave background radiation, rest on the violation of the law of conservation of energy by the expanding space. However physics of "space creation" can not be tested in laboratory and hence needs more observational evidences.

The most perspective and clear observational test on the reality of the space expansion was suggested by Sandage (1962) and related to the measurement of the variation of redshift with time ($\dot{z} = dz/dt$), which is within the reach by the OWL ESO telescope (Pasquini et al. 2005).

The big bang SCM is not the ultimate model of the Universe. There are several cosmological models which are based on other fundamental hypotheses and give different interpretation of observable phenomena. A classification of possible relativistic cosmologies in accordance with basic initial assumptions were discussed by Baryshev et al.(1994) [2]. In particular, relativistic quantum field approach to gravity (Baryshev

2008a) where the Minkowski space and conservation laws are valid, may be considered as the basis of the field gravity fractal cosmological model (Baryshev 2008b). Crucial observational tests of alternative cosmological models and gravity theories should be developed to get a progress in the cosmological physics.